\documentclass{webofc}
\usepackage[varg]{txfonts}   
\begin{document}
\title{Collectivity in ultra-peripheral heavy-ion collisions}
%
%

\author{\firstname{Chun} \lastname{Shen}\inst{1,2}\fnsep\thanks{\email{chunshen@wayne.edu}} \and
        \firstname{Wenbin} \lastname{Zhao}\inst{1} \and
        \firstname{Bj\"orn} \lastname{Schenke}\inst{3}
}

\institute{Department of Physics and Astronomy, Wayne State University, Detroit, Michigan 48201, USA
\and
           RIKEN BNL Research Center, Brookhaven National Laboratory, Upton, NY 11973, USA
\and
           Physics Department, Brookhaven National Laboratory, Upton, NY 11973, USA
          }

\abstract{%
  We present full (3+1)D dynamical simulations to study collective behavior in ultra-peripheral nucleus-nucleus collisions (UPC) at the Large Hadron Collider (LHC) with the 3DGlauber+MUSIC+UrQMD framework \cite{Zhao:2022ayk, Shen:2022oyg}. By extrapolating from asymmetric p+Pb collisions, we simulate a quasi-real photon $\gamma^*$ interacting with the Pb nucleus in an ultra-peripheral collision at the LHC, assuming strong final-state effects. We study the elliptic flow hierarchy between p+Pb and $\gamma^*$+Pb collisions, which is dominated by the difference in longitudinal flow decorrelations.
  Our theoretical framework provides a quantitative tool to study collectivity in small asymmetric collision systems at current and future collider experiments.
}
\maketitle

\section{Introduction}
\label{intro}

Collective features of strongly-coupled systems have been observed in relativistic nuclear collisions with light and heavy nuclei, such as p+Au, d+Au, $^3$He+Au at the Relativistic Heavy-Ion Collider (RHIC)~\cite{PHENIX:2017xrm,PHENIX:2018lia}, and p+p and p+Pb collisions at the Large Hadron Collider (LHC)~\cite{Li:2012hc,Dusling:2015gta,Nagle:2018nvi}. The theoretical interpretation of these flow-like signals has been a hot topic, driving our field to unravel how the collective behavior emerges depending on the collision system size~\cite{Shen:2020mgh, Schenke:2021mxx}.
Recently, the ATLAS Collaboration measured the two-particle azimuthal correlations in ultra-peripheral Pb+Pb collisions (UPCs) at the LHC~\cite{ATLAS:2021jhn}. The high multiplicity UPC events created from the photo-nuclear interactions showed the persistence of collective phenomena with correlations comparable to those observed in p+p and p+Pb collisions at similar multiplicity~\cite{ATLAS:2021jhn}.

Quantitative understanding of the many-body dynamics in these small collision systems requires the development and application of full (3+1)D simulations beyond Bjorken's boost-invariance paradigm in the high energy limit~\cite{Bozek:2015swa,Ke:2016jrd,Schenke:2016ksl,Shen:2017bsr,Shen:2020jwv,Wu:2021hkv,Schafer:2021csj,Lisa:2021zkj}. In photon-nucleus collisions, the quasi-real photon $\gamma^*$'s energy fluctuates event-by-event, and is much smaller than the energy of the incoming Pb nucleus. Such unbalanced and fluctuating kinematics leads to a highly asymmetric collision system, strongly violating the longitudinal boost invariance. In these asymmetric systems, the rapidity decorrelation of the collision geometry plays a crucial role when computing and measuring the anisotropic flow coefficients.

In this proceeding, we study the flow rapidity decorrelation in detail for $\gamma^*$+Pb and p+Pb collisions, providing complementary information to Ref.~\cite{Zhao:2022ayk}.

\section{Fluctuations in collision kinematics of $\gamma^*$+A collisions in UPCs}
\label{sec-2}

The fast-moving Pb spectators in the UPC events generate strong fluxes of quasi-real photons. The emitted photons have the following energy  spectrum~\cite{Bertulani:2005ru,Baltz:2007kq},
\begin{equation}
    \frac{dN^\gamma}{dk_\gamma} = \frac{2 Z^2 \alpha}{\pi k_\gamma}\left[ w_R^{AA} K_0(w_R^{AA}) K_1(w_R^{AA}) - \frac{(w_R^{AA})^2}{2}(K_1^2(w_R^{AA}) - K_0^2(w_R^{AA})) \right],
    \label{eq:1}
\end{equation}
where $\alpha = 1/137$ and $w_R^{AA} = 2 k_\gamma R_A/\gamma_L$, with the longitudinal Lorentz contraction factor $\gamma_L = \sqrt{s_\mathrm{NN}}/(2m_N)$. The functions $K_0(x)$ and $K_1(x)$ are the modified Bessel functions of the second kind. For the Pb nucleus, $R_A = 6.62$\,fm and $Z = 82$. The kinematics for incoming photon projectile and nucleon target in the Pb nucleus are $P^\mu_\gamma  \simeq (k_\gamma, 0, 0, k_\gamma)$ and $P^\mu_N \simeq (\sqrt{s_{\rm NN}}/2, 0, 0, -\sqrt{s_{\rm NN}}/2)$, where we neglect the photon's virtuality and nucleon's rest mass. The center of mass collision energy for the $\gamma^*$+A system is
\begin{equation}
    \sqrt{s_{\gamma^* N}} = \sqrt{2 k_\gamma \sqrt{s_{\rm NN}}}.
    \label{eq:2}
\end{equation}
From Eqs.~(\ref{eq:1}) and (\ref{eq:2}), we can compute the probability distribution for the center of mass energy in $\gamma^*$+A collisions,
\begin{equation}
    P(\sqrt{s_{\gamma^*N}}) \propto \frac{\sqrt{s_{\gamma^* N}}}{k_\gamma}\left[ w_R^{AA} K_0(w_R^{AA}) K_1(w_R^{AA}) - \frac{(w_R^{AA})^2}{2}(K_1^2(w_R^{AA}) - K_0^2(w_R^{AA})) \right],
    \label{eq:3}
\end{equation}
with the photon momentum $k_\gamma = s_{\gamma^* N}/(2\sqrt{s_{\rm NN}})$.
Because of the unequal incoming longitudinal momentum between the quasi-real photons and the target nucleon, the center of mass rapidity of the $\gamma^*$+Pb system differs from the lab frame rapidity by 
\begin{equation}
    \Delta y = y_{\rm beam}(\sqrt{s_{\gamma^* N}}) - y_{\rm beam}(\sqrt{s_{\rm NN}}),
\end{equation}
where the beam rapidity for a given center-of-mass collision energy can be computed as $y_{\rm beam}(\sqrt{s}) = {\rm arccosh}(\sqrt{s}/(2 m_N))$.

\begin{figure}[b!]
    \centering
    \includegraphics[width=0.48\linewidth]{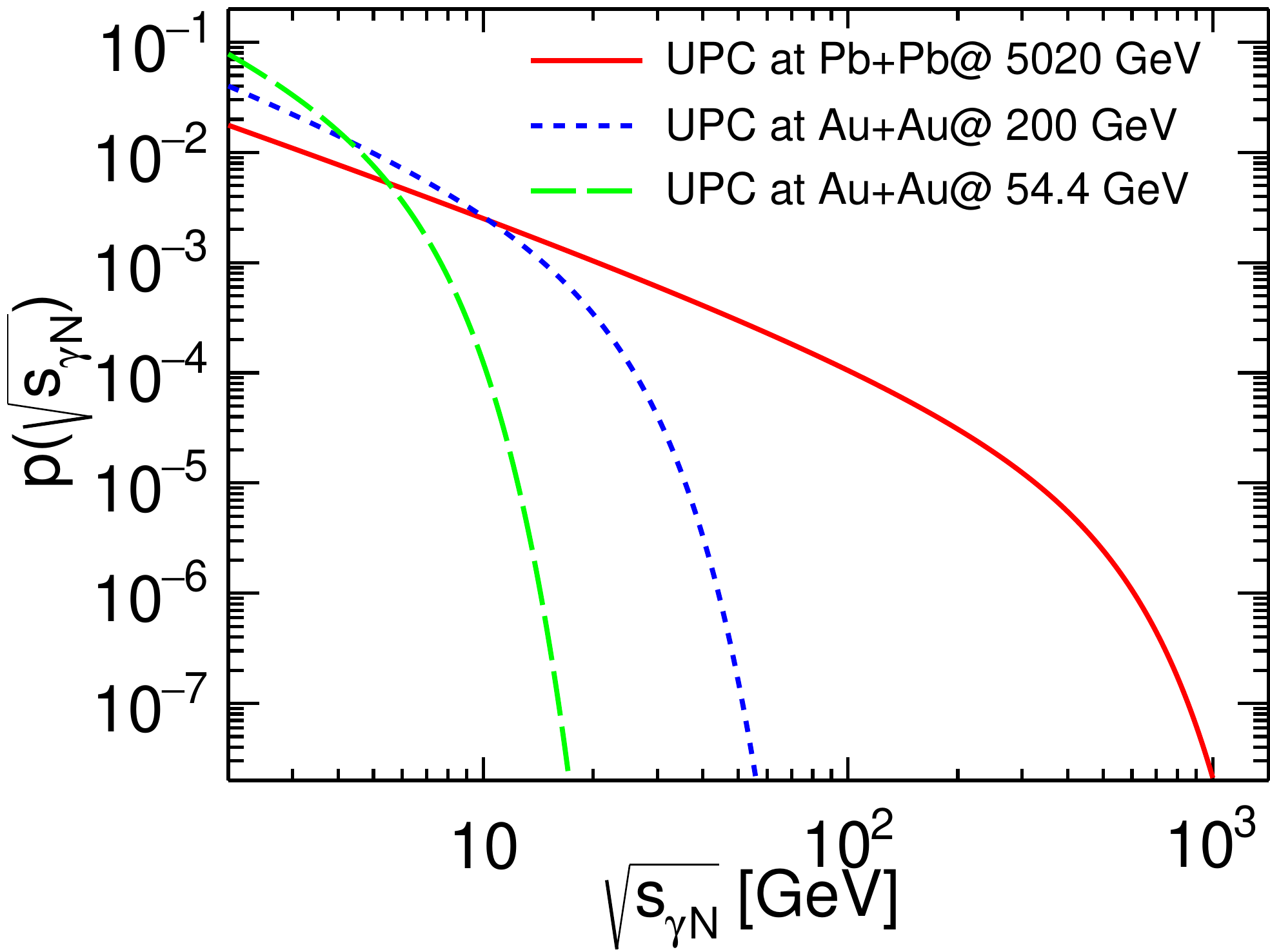}
    \includegraphics[width=0.48\linewidth]{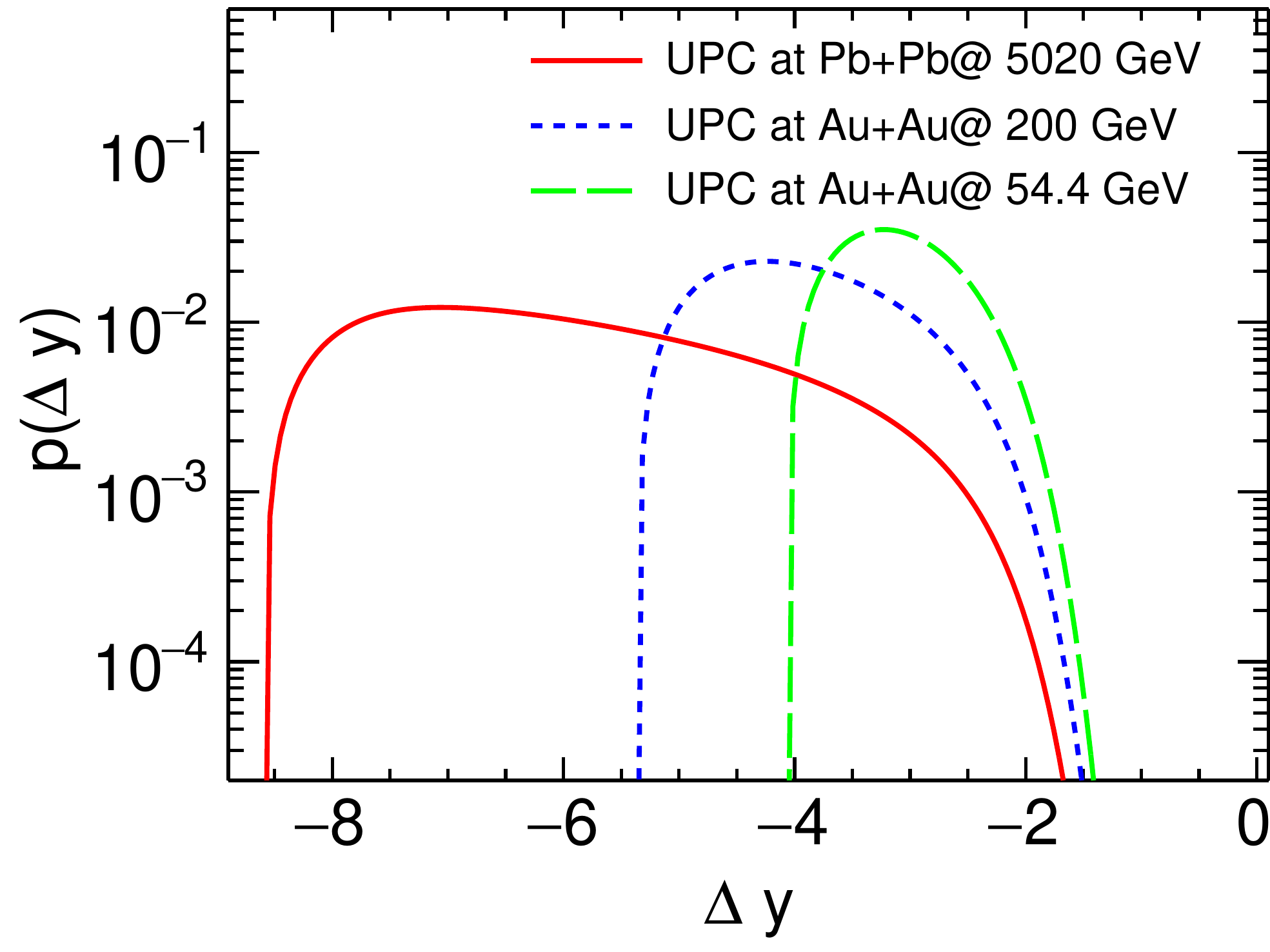}
    \caption{\textit{Left Panel:} The probability distributions of the center-of-mass collision energies for photon-nucleus collisions in Au+Au and Pb+Pb UPC events at three collision energies. \textit{Right Panel:} The probability distributions of the global rapidity shifts in $\gamma^*$+A collisions from the center-of-mass frame to the lab frame. Negative $\Delta y$ represents the shift towards the nucleus-going direction. }
    \label{fig:1}
\end{figure}
The left panel of Figure~\ref{fig:1} shows the probability distributions of the center-of-mass collision energies in Au+Au and Pb+Pb UPC events at RHIC and LHC. 
The center-of-mass collision energies in $\gamma^*$+A collisions are much smaller than in their corresponding heavy-ion collisions. The values of $\sqrt{s_{\gamma^* N}}$ fluctuate over wide ranges, which results in broad intervals for rapidity shifts between the center-of-mass frame and lab frame, as shown in the right panel of Figure~\ref{fig:1}. For UPC events in Pb+Pb collisions at 5020 GeV, the rapidity shifts fluctuate from $-2$ to $-8.5$.
Therefore, it is important to include these kinematics fluctuations in $\gamma^*$+A collisions, which result in non-trivial effects in the rapidity direction. We note that small collision energy and large global rapidity shift result in little particle production at mid-rapidity in the lab frame. Therefore, triggering high multiplicity events at mid-rapidity effectively selects the $\gamma^*$+A collisions with large $\sqrt{s_{\gamma^* N}}$.

\section{Flow rapidity decorrelation in $\gamma^*$+Pb and p+Pb collisions}
\label{sec-3}

In the work~\cite{Zhao:2022ayk}, we found the different amounts of longitudinal flow decorrelations in $\gamma^*$+Pb and p+Pb collisions led to the elliptic flow hierarchy observed by the ATLAS Collaboration~\cite{ATLAS:2021jhn}. The different flow rapidity decorrelations in $\gamma^*$+Pb and p+Pb collisions come from the difference in center-of-mass collision energy and the global rapidity shift in $\gamma^*$+Pb collisions~\cite{Zhao:2022ayk}.

\begin{figure}[h!]
    \centering
    \includegraphics[width=0.48\linewidth]{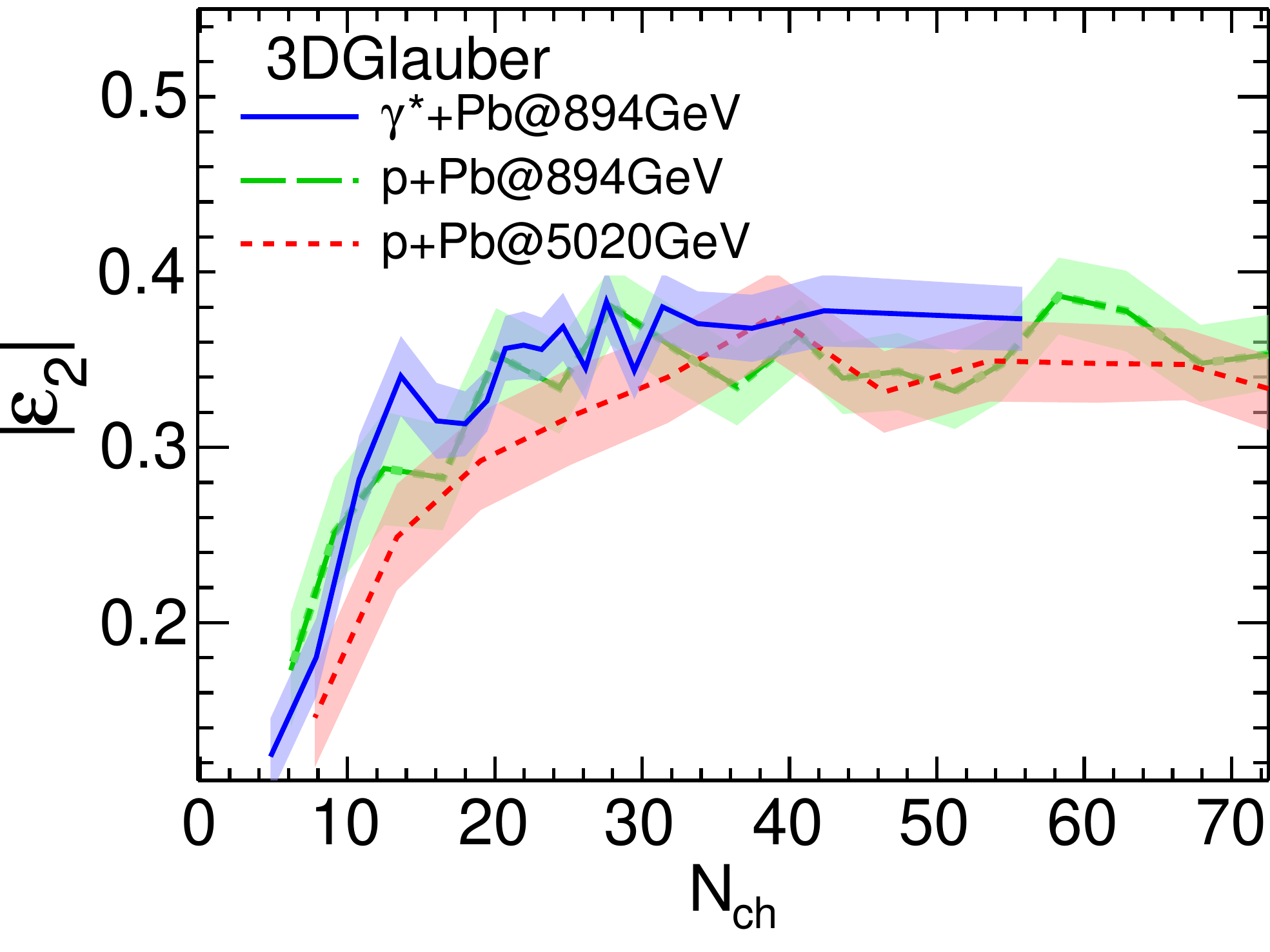}
    \includegraphics[width=0.48\linewidth]{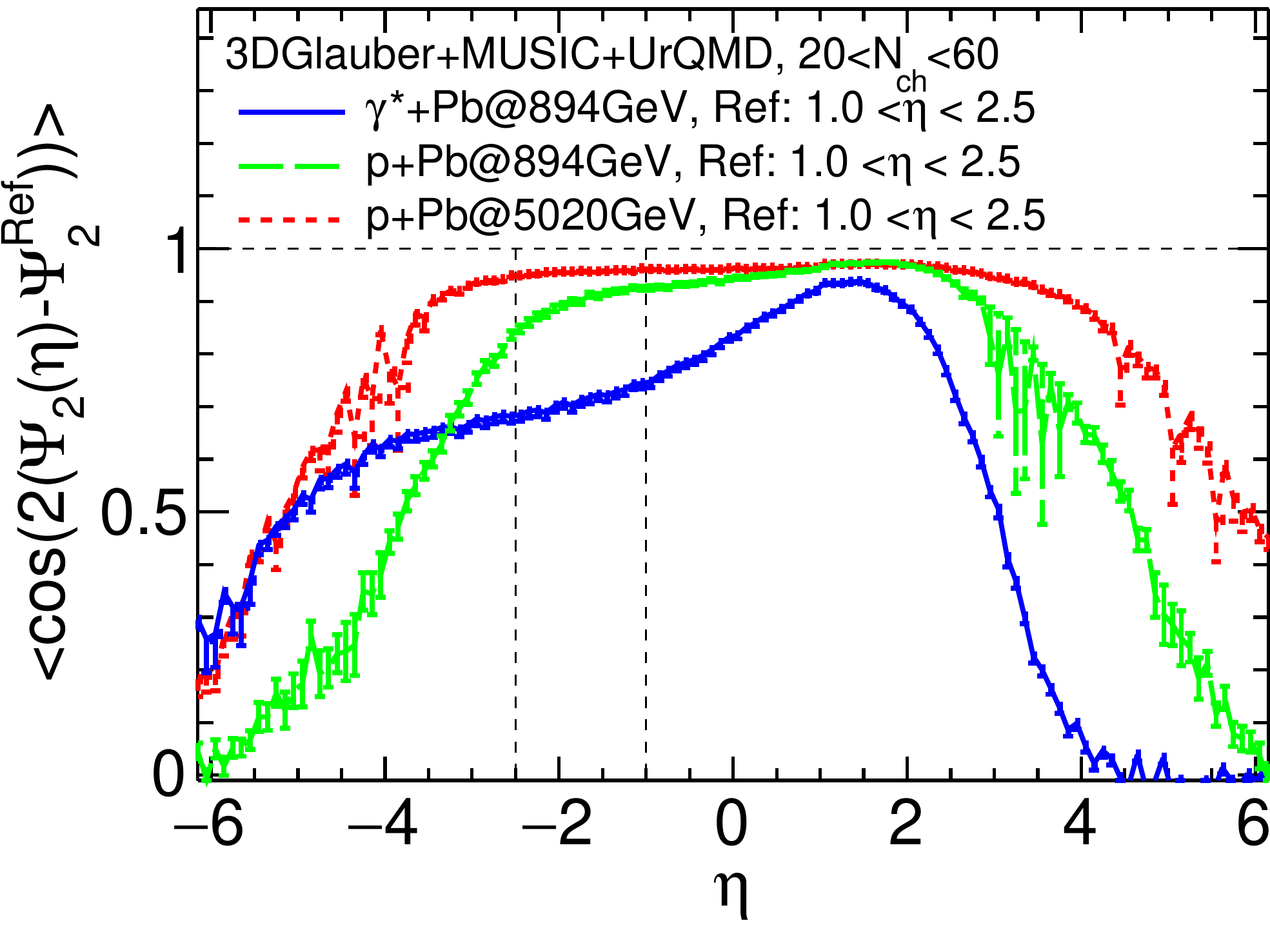}
    \caption{\textit{Left Panel:} Initial-state ellipticities of the fireballs in $1 < |\eta_s| < 2.5$ at $\tau = 0.6$\,fm/c for $\gamma^*$+Pb and p+Pb collisions at $\sqrt{s} = 894$\,GeV and $\sqrt{s} = 5020$\,GeV. \textit{Right Panel:} Final-state elliptic flow event-plane correlation with respect to a reference flow angle $\psi_2^{\rm ref}$ at $1 < \eta < 2.5$ for $\gamma^*$+Pb and p+Pb collisions at two collision energies.}
    \label{fig:2}
\end{figure}

To investigate the flow rapidity correlations at a charged hadron multiplicity of $40-50$, we focus on the following analysis with $\gamma^*$+Pb collisions at their highest energy, 894 GeV. The left panel of Figure~\ref{fig:2} shows that the values of initial-state ellipticity are almost the same between $\gamma^*$+Pb and p+Pb collisions as a function of the particle multiplicity, which means that the shape fluctuations in the transverse plane are at the same level for the two collision systems.
The right panel of Figure~\ref{fig:2} shows the evolution of event-plane correlations in steps from p+Pb collisions at 5020 GeV to $\gamma^*$+Pb at 894 GeV collisions. The ATLAS Collaboration measured the two-particle correlation with a rapidity gap of $| \Delta \eta | > 2$. This analysis method computes the flow angular correlation between the two pseudorapidity intervals, namely $\eta_1 \in [-2.5, -1]$ and $\eta_2 \in [1, 2.5]$. The right panel of Figure~\ref{fig:2} shows that the angular correlations of the elliptic flow vectors are strong in p+Pb collisions at 5020 GeV, very close to unity between these two $\eta$ intervals. Reducing the collision energy to 894 GeV shortens the length of the produced strings in the rapidity space, weakening the event-plane correlation to $\sim0.9$. The extra global rapidity shift in $\gamma^*$+Pb collisions further reduces the correlation strength to $\sim 0.7$. 

\section{Conclusion}
\label{sec-4}

We applied a newly developed (3+1)D dynamical framework to study the collectivity in highly asymmetric relativistic nuclear collisions, such as p+A collisions and $\gamma^*$+A in the ultra-peripheral A+A collisions at RHIC and LHC energies~\cite{Zhao:2022ayk, Shen:2022oyg}.
In this proceeding, we present a detailed analysis of the flow rapidity decorrelation in $\gamma^*$+Pb and p+Pb collisions. 
We discuss how to include fluctuating collision energies in simulating the 3D dynamics of $\gamma^*$+A collisions in the UPC events. At LHC energies, the elliptic flow hierarchy between the $\gamma^*$+Pb and p+Pb collisions can be explained by the different amounts of flow rapidity decorrelations in these systems, demonstrating the necessity of full (3+1)D simulations for these asymmetric collision systems.

\section*{Acknowledgements}
W.B.Z. is supported by the National Science Foundation (NSF) under grant numbers ACI-2004571 within the framework of the XSCAPE project of the JETSCAPE collaboration. B.P.S. and C.S. are supported by the U.S. Department of Energy, Office of Science, Office of Nuclear Physics, under DOE Contract No.\,DE-SC0012704, and Award No. DE-SC0021969, respectively.
This research was done using resources provided by the Open Science Grid (OSG) \cite{Pordes:2007zzb, Sfiligoi:2009cct}, which is supported by the National Science Foundation award \#2030508.

\bibliography{references}

\begin{thebibliography}{22}

\bibitem{Zhao:2022ayk}
W.~Zhao, C.~Shen, B.~Schenke (2022), \texttt{2203.06094}

\bibitem{Shen:2022oyg}
C.~Shen, B.~Schenke, Phys. Rev. C \textbf{105}, 064905 (2022),
  \texttt{2203.04685}

\bibitem{PHENIX:2017xrm}
C.~Aidala et~al. (PHENIX), Phys. Rev. Lett. \textbf{120}, 062302 (2018),
  \texttt{1707.06108}

\bibitem{PHENIX:2018lia}
C.~Aidala et~al. (PHENIX), Nature Phys. \textbf{15}, 214 (2019),
  \texttt{1805.02973}

\bibitem{Li:2012hc}
W.~Li, Mod. Phys. Lett. A \textbf{27}, 1230018 (2012), \texttt{1206.0148}

\bibitem{Dusling:2015gta}
K.~Dusling, W.~Li, B.~Schenke, Int. J. Mod. Phys. E \textbf{25}, 1630002
  (2016), \texttt{1509.07939}

\bibitem{Nagle:2018nvi}
J.L. Nagle, W.A. Zajc, Ann. Rev. Nucl. Part. Sci. \textbf{68}, 211 (2018),
  \texttt{1801.03477}

\bibitem{Shen:2020mgh}
C.~Shen, L.~Yan, Nucl. Sci. Tech. \textbf{31}, 122 (2020), \texttt{2010.12377}

\bibitem{Schenke:2021mxx}
B.~Schenke, Rept. Prog. Phys. \textbf{84}, 082301 (2021), \texttt{2102.11189}

\bibitem{ATLAS:2021jhn}
G.~Aad et~al. (ATLAS), Phys. Rev. C \textbf{104}, 014903 (2021),
  \texttt{2101.10771}

\bibitem{Bozek:2015swa}
P.~Bozek, A.~Bzdak, G.L. Ma, Phys. Lett. B \textbf{748}, 301 (2015),
  \texttt{1503.03655}

\bibitem{Ke:2016jrd}
W.~Ke, J.S. Moreland, J.E. Bernhard, S.A. Bass, Phys. Rev. C \textbf{96},
  044912 (2017), \texttt{1610.08490}

\bibitem{Schenke:2016ksl}
B.~Schenke, S.~Schlichting, Phys. Rev. C \textbf{94}, 044907 (2016),
  \texttt{1605.07158}

\bibitem{Shen:2017bsr}
C.~Shen, B.~Schenke, Phys. Rev. C \textbf{97}, 024907 (2018),
  \texttt{1710.00881}

\bibitem{Shen:2020jwv}
C.~Shen, S.~Alzhrani, Phys. Rev. C \textbf{102}, 014909 (2020),
  \texttt{2003.05852}

\bibitem{Wu:2021hkv}
X.Y. Wu, G.Y. Qin (2021), \texttt{2109.03512}

\bibitem{Schafer:2021csj}
A.~Sch\"afer, I.~Karpenko, X.Y. Wu, J.~Hammelmann, H.~Elfner (2021),
  \texttt{2112.08724}

\bibitem{Lisa:2021zkj}
M.A. Lisa, J.a.G.P. Barbon, D.D. Chinellato, W.M. Serenone, C.~Shen,
  J.~Takahashi, G.~Torrieri, Phys. Rev. C \textbf{104}, 011901 (2021),
  \texttt{2101.10872}

\bibitem{Bertulani:2005ru}
C.A. Bertulani, S.R. Klein, J.~Nystrand, Ann. Rev. Nucl. Part. Sci.
  \textbf{55}, 271 (2005), \texttt{nucl-ex/0502005}

\bibitem{Baltz:2007kq}
A.J. Baltz, Phys. Rept. \textbf{458}, 1 (2008), \texttt{0706.3356}

\bibitem{Pordes:2007zzb}
R.~Pordes et~al., J. Phys. Conf. Ser. \textbf{78}, 012057 (2007)

\bibitem{Sfiligoi:2009cct}
I.~Sfiligoi, D.C. Bradley, B.~Holzman, P.~Mhashilkar, S.~Padhi, F.~Wurthwrin,
  WRI World Congress \textbf{2}, 428 (2009)

\end{thebibliography}

\end{document}